\documentclass{IEEEtran} 
\usepackage{amsmath,amssymb,verbatim,rotating} 

\newtheorem{lemma}{Lemma}

\newcommand{\mycite}[1]{~\cite{#1}}

\newcommand{\mymathbb}[1]{{\mathbb{#1}}} 
\newcommand{\mymathsf}[1]{{\mathsf{#1}}} 

\newcommand{\dmn}{q} 
\newcommand{\cA}{{\cal A}}
\newcommand{\sA}{\mymathsf{A}}
\newcommand{\cB}{{\cal B}}

\newcommand{\bC}{\mymathbb{C}}

\newcommand{\cE}{{\cal E}}
\newcommand{\sE}{\mymathsf{E}}
\newcommand{\cF}{{\cal F}}

\newcommand{\myF}{{\mymathbb{F}_{\dmn}}} 
\newcommand{\myFnoarg}{{\mymathbb{F}}}

\newcommand{\cN}{{\cal N}}

\newcommand{\sH}{\mymathsf{H}} 

\newcommand{\sL}{\mymathsf{L}}

\newcommand{\sQ}{\mymathsf{Q}}
\newcommand{\cQ}{{\cal Q}}

\newcommand{\cR}{{\cal R}}

\newcommand{\sV}{\mymathsf{V}}

\newcommand{\cX}{{\cal X}}

\newcommand{\sY}{\mymathsf{Y}}

\newcommand{\cY}{{\cal X}} 

\newcommand{\wbX}[1]{{X}^{#1}} 
\newcommand{\wbZ}[1]{{Z}^{#1}}

\renewcommand{\subset}{\subseteq}
\renewcommand{\tilde}{\widetilde}
\renewcommand{\hat}{\widehat}

\newcommand{\mbm}[1]{\mbox{\boldmath $#1$}}

\newcommand{\cmple}{^{\rm c}}

\newcommand{\SINT}{\mymathbb{Z}}

\newcommand{\Expe}{\mymathbb{E}} 

\newcommand{\Prob}{{\rm Pr}}

\newcommand{\tnsr}{\otimes}
\newcommand{\trace}{{\rm Tr}\,} 

\newcommand{\lag}{\langle}
\newcommand{\rag}{\rangle}
\newcommand{\crd}[1]{|#1|}
\newcommand{\bra}[1]{\lag #1 |}
\newcommand{\ket}[1]{| #1 \rag}

\newcommand{\sypnoarg}{\mymathsf{f}_{\rm sp}}
\newcommand{\syp}[2]{\sypnoarg(#1,  #2)}
\newcommand{\dpr}[2]{#1 \cdot #2}

\newcommand{\perpsyp}{\perp_{\rm sp}}

\newcommand{\Hch}{{\sH}}

\newcommand{\Hgen}{{\sH^{\tnsr n}}}

\newcommand{\Bop}{\sL} 

\newcommand{\Fen}{F_{\rm e}}

\newcommand{\imu}{{\rm i}} 
\newcommand{\Ebe}{N}

\newcommand{\ketbe}[1]{\ket{#1}}
\newcommand{\phasebe}{\omega}
\newcommand{\Xbe}{X}
\newcommand{\Zbe}{Z}

\newcommand{\rvX}{\mymathsf{X}}

\newcommand{\rvy}{\mymathsf{Y}}

\newcommand{\varn}{n}   

\newcommand{\rvZ}{\mymathsf{Z}} 

\newcommand{\rvx}{\rvX} 
\newcommand{\rvz}{\rvZ} 
\newcommand{\rvv}{\sV}

\newcommand{\rve}{\sE}

\newcommand{\Ccl}{C}

\newcommand{\Se}[2]{S_{#1 #2}}

\newcommand{\myFpower}[1]{\mymathbb{F}_{\dmn}^{#1}}

\newcommand{\Qpl}[1]{\sQ^+} 
\newcommand{\Qmi}[1]{\sQ^-}

\newcommand{\Bp}[1]{G}

\newcommand{\Jgood}{\tilde{J}}

\newcommand{\Ksp}{J} 
\newcommand{\crI}{J} 

\newcommand{\Jof}[1]{\prm(\Jgood)} 
\newcommand{\Kof}[1]{\Ksp(#1)}

\newcommand{\prm}{\pi}

\newcommand{\rvxi}{\mbm{\xi}} 
\newcommand{\rvzeta}{\mbm{\zeta}} 

\newcommand{\Cite}[1]{\cite{#1}}

\newcommand{\crJ}{\Gamma}
\newcommand{\CSone}{C_1}
\newcommand{\CStwo}{C_2^{\perp}} 
\newcommand{\CStwp}{C_2}

\newcommand{\varg}{y} 
\newcommand{\soE}{\cF} 
\newcommand{\subgrp}{\le}

\title{Conjugate Codes and Applications\\ to Cryptography}

\author{%
  Mitsuru Hamada\\[1ex]
Research Center for Quantum Information Science\\
   Tamagawa University Research Institute\\
6-1-1 Tamagawa-gakuen, Machida, Tokyo 194-8610, Japan\\[0.7ex]
PRESTO, Japan Science and Technology Agency\\ 
4-1-8 Honcho, Kawaguchi, Saitama, Japan
}  

\begin{document}

\maketitle

\begin{abstract}
A conjugate code pair is defined as a pair of linear codes
such that one contains the dual of the other.
The conjugate code pair represents the essential structure
of the corresponding Calderbank-Shor-Steane (CSS) quantum code. 
It is argued that conjugate code pairs are applicable to
quantum cryptography in order to motivate studies on 
conjugate code pairs.
\end{abstract}


\section{Introduction \label{ss:intro}}

Since the invention of the first algebraic quantum error-correcting
code (QECC) by Shor\mycite{shor95} in 1995, 
the theory of QECCs has been developed rapidly. 
The first code was soon extended to a class of algebraic QECCs
called Calderbank-Shor-Steane (CSS) codes~\cite{CalderbankShor96} and then to
a more general class of QECCs, which are
called symplectic codes or stabilizer
codes\mycite{crss97,crss98,gottesman96}. 

In this paper, we focus on CSS codes.
It is well-known that this class of symplectic codes 
are useful for quantum key distribution (QKD), at least, in theory.
In particular, Shor and Preskill\mycite{ShorPreskill00}
argued that the security of the famous 
Bennett-Brassard 1984 (BB84) QKD protocol 
could be proved 
by evaluating the fidelity of quantum error-correcting codes underlying
the protocol. 

The term `conjugate codes' appearing in the title is
almost a synonym for CSS codes if one forgets about quantum mechanical
operations for encoding or decoding and pays attention only to
what can be done in the coding theorists' universe of finite fields.
This term was coined here so that this issue would be more accessible
to those unfamiliar with quantum information theory. 

Recently, the present author~\cite{hamada03s} proved the existence of CSS codes
that outperforms those proved to exist 
in the literature~\cite{CalderbankShor96}, and 
quantified the security and reliability of CSS-code-based QKD 
schemes rigorously assuming ideal discrete quantum systems. 
Although we have treated QKD in \cite{hamada03s}, the 
CSS-code-based QKD scheme
can be viewed as merely one application of conjugate codes (CSS codes).
For example,
the arguments in \cite{hamada03s} also imply that 
conjugate codes can be used as cryptographic codes that directly encrypt secret data
as will be elucidated in the sequel.
Here, we remark that QKD means techniques 
for sharing a secret key between remote parties,
and the shared key itself is not the secret message that 
the sender wishes to send. 
A typical scenario is that after sharing the key, the sender encrypts
a secret data using the key and sends it to the receiver and
the receiver decrypts the data using the shared key.
The direct encryption is mightier, 
and can be used as QKD if one wishes.
Turning back to the original motivation of (algebraic) QECCs, these codes 
deemed indispensable for quantum computing
since quantum states are more vulnerable to errors or quantum noise, 
most notably to decoherence.
Among QECCs, CSS codes are said to be suited for fault-tolerant quantum computing
(e.g., \cite{steane99} and references therein). 

The aim of this work is to enhance motivation to study
this class of codes. In particular, applications
to cryptography, which allow direct encryption, are emphasized. 
We remark that a large portion of this paper is nearly a paraphrase of a part of
\cite{hamada03s} though our description is slightly more general
in that we explicitly treat a general code pair $(C_1,C_2)$ satisfying 
a certain condition, which will be given shortly,
whereas \cite{hamada03s} describes the result for the case where $C_1=C_2$.

This paper is organized as follows.
In Section~\ref{ss:cc}, conjugate codes are introduced, and
in Section~\ref{ss:css}, CSS quantum codes are explained.
In Section~\ref{ss:css_crypt}, 
it is argued that conjugate code pairs are applicable to
quantum cryptography.
General symplectic codes, and quotient codes, are explained in 
Sections~\ref{ss:gsc} and \ref{ss:qc}, respectively.
Sections~\ref{ss:rmk} and \ref{ss:cnc}, 
contain remarks and a summary, respectively.

\section{Conjugate Codes \label{ss:cc}}

We write $B \subgrp C$
if $B$ is a subgroup of an additive group $C$.
We use a finite field $\myF$ of $q$ elements, and the dot product defined by
\begin{equation}\label{eq:dotpr}
\dpr{(x_1,\dots,x_n)}{(y_1,\dots,y_n)}=\sum_{i=1}^{n} x_iy_i
\end{equation}
for vectors in $\myFpower{n}$.
We let $C^{\perp}$ denote $\{ y \in\myFpower{n} \mid \forall x\in C, \ \dpr{x}{y}=0 \}$
for a 
subset $\Ccl$ of $\myFpower{n}$.

We mean by an $[[n,k]]$ {\em conjugate (complementary) code pair}\/
or CSS code pair over $\myF$
a pair $(C_1,C_2)$ consisting of
an $[n,k_1]$ linear code $\CSone$ and an
$[n,k_2]$ linear code $C_2$ 
satisfying%
\footnote{%
When the 
number of elements of a code $C \subset\myFpower{n}$ is
$\dmn^k$, it is called an $[n,k]$ code. Readers unfamiliar with coding theory
are referred to \cite[Section~2]{hamada05qc} or standard textbooks
such as~\cite{mceliece,peterson,vanLint3rd,sloan,berlekampkp}.%
}
\begin{equation}\label{eq:css_cond}
\CStwo \subgrp \CSone, 
\end{equation}
which condition is equivalent to
$\CSone^{\perp} \subgrp \CStwp$,
and
\begin{equation}
k=k_1+k_2-n.
\end{equation}
If $C_1$ and $C_2$ satisfy (\ref{eq:css_cond}), 
the quotient codes $C_1/C_2^{\perp}$ and $C_2/C_1^{\perp}$ 
are said to be conjugate.
The notion of quotient codes was introduced in \cite{hamada05qc}, 
and will be explained in Section~\ref{ss:qc}.

The goal, in a long span, 
is to find a conjugate code pair $(C_1,C_2)$ such that
both $C_1/C_2^{\perp}$ and $C_2/C_1^{\perp}$ 
have good performance. If the linear codes $C_1$ and $C_2$ both have
good performance, so do $C_1/C_2^{\perp}$ and $C_2/C_1^{\perp}$. 
Hence, a conjugate code pair $(C_1,C_2)$ with good (not necessarily a technical term) 
$C_1$ and $C_2$ is also desirable.


\section{Calderbank-Shor-Steane Codes \label{ss:css}}

The complex linear space of operators on a Hilbert space $\Hch$ is
denoted by $\Bop(\Hch)$.
A quantum code usually means a pair $(\cQ,\cR)$ consisting of a subspace $\cQ$
of $\Hch^{\tnsr n}$ and a trace-preserving completely positive
(TPCP) linear map $\cR$ on
$\Bop(\Hch^{\tnsr n})$, called a recovery operator. The subspace $\cQ$ alone
is also called a code. 
Symplectic codes have more structure: They are
simultaneous eigenspaces of commuting operators on
$\Hch^{\tnsr n}$.
Once a set of commuting operators is specified,
we have a collection of eigenspaces of them.
A symplectic code refers to either such an eigenspace or a collection 
of eigenspaces, 
each possibly accompanied by a suitable recovery operator.
In this section and the next, 
we assume $\Hch$ is a Hilbert space of dimension $\dmn$, and
$\dmn$ is a prime (but see Section~\ref{ss:npa}). 
Then, $\myF=\SINT/\dmn\SINT$. We fix an orthonormal basis $(\ket{i})_{i=0}^{\dmn-1}$ of $\Hch$.

In constructing symplectic codes, 
the following basis of $\Bop(\Hch^{\tnsr n})$ is used.
Let unitary operators $\Xbe, \Zbe$ on $\Hch$ 
be defined by
\begin{equation}\label{eq:error_basis}
\Xbe \ketbe{j}  = \ketbe{j-1}, \,\,\,
\Zbe \ketbe{j} = \phasebe^ j \ketbe{j}, \quad \,\,\, j\in\myF
\end{equation}
with $\phasebe$ being a primitive $\dmn$-th root of unity (e.g., $e^{\imu 2\pi/\dmn}$).
For $u=(u_1,\dots,u_n)\in\myFpower{n}$, let $\wbX{u}$ and $\wbZ{u}$ denote
$X^{u_1}\tnsr \cdots \tnsr X^{u_n}$
and $Z^{u_1}\tnsr \cdots \tnsr Z^{u_n}$, 
respectively.
The operators $\wbX{u}\wbZ{w}$, $u,w\in \myFpower{n}$,
form a basis of 
$\Bop(\Hch^{\tnsr n})$, which we call the Weyl (unitary)
basis~\cite{weyl28}.
We have the commutation relation
\begin{equation}\label{eq:WCR}
(X^{u}Z^{w})(X^{u'}Z^{w'})=\omega^{\dpr{u}{w'}-\dpr{w}{u'}}
(X^{u'}Z^{w'})(X^{u}Z^{w}), 
\end{equation}
for $u,w,u',w'\in\myFpower{n}$, 
which follows from $XZ=\omega ZX$.
It is sometimes useful to rearrange the components of $(u,w)$ 
appearing in the operators $\wbX{u}\wbZ{w}$ in the Weyl basis as follows:
For $u=(u_1,\dots,u_n)$ and $w=(w_1,\dots,w_n) \in\myFpower{n}$,
we denote by $[u,w]$ the rearranged one 
\[
\big((u_1,w_1),\dots,(u_n,w_n)\big) \in \cX^n, 
\]
where $\cX=\myF\times\myF$.
We occasionally use another symbol $\Ebe$ for the Weyl basis:
\[
\Ebe_{[u,w]}=\wbX{u}\wbZ{w}
\]
and
\[
\Ebe_{\Ksp}= \{ \Ebe_x \mid x\in \Ksp \}, \quad \Ksp\subset\cX^n.
\]
An obvious but important consequence of (\ref{eq:WCR})
is that $X^{u}Z^{w}$ and $X^{u'}Z^{w'}$ commute if and only if
$\dpr{u}{w'}-\dpr{w}{u'}=0$. The map 
\begin{equation}\label{eq:syp_blf}
([u,w],[u',w'])\mapsto\dpr{u}{w'}-\dpr{w}{u'}
\end{equation}
is a symplectic bilinear form,
which we refer to as standard.

A CSS code is specified by two classical linear 
codes (i.e., subspaces of $\myFpower{n}$) 
$\CSone$ and $\CStwp$ with (\ref{eq:css_cond}).%
\footnote{Our code pair $(C_1, C_2)$ often appears
as $(C_1,C_2^{\perp})$ in the literature~\cite{CalderbankShor96,ShorPreskill00,hamada05qc}. Our choice would be more acceptable to coding theorists because 
good (not necessarily a technical term) codes $C_1$ and $C_2$ result in a good CSS code while the performance of 
$C_2^{\perp}$ seemingly has no direct meaning.}
Coset structures are exploited in construction of CSS codes.
We fix some set of coset representatives
of the factor group $\myFpower{n}/\CSone$, for which the             
letter $x$ is always used to refer to a coset representative
in this section,
that of $\CSone/\CStwo$, for which $v$ is used,
and that of $\myFpower{n}/\CStwp$, for which $z$ is used.
These may be written as
\begin{eqnarray*}
x &\in & \myFpower{n}/\CSone,\\
z &\in & \myFpower{n}/\CStwp,\\
v &\in & \CSone/\CStwo
\end{eqnarray*}
where $\in$ is in abuse as usual.
Put $k_1=\dim \CSone$, $k_2=\dim \CStwp$, 
and assume $g_1,\dots, g_{n-k_2}$ form a basis of $\CStwo$,
and $h_1,\dots, h_{n-k_1}$ form a basis of $\CSone^{\perp}$.
We assume $k_1$ is 
larger than $n-k_2$.

The operators
\begin{equation}\label{eq:stab4CSS}
 \wbZ{h_1}, \dots, \wbZ{h_{n-k_1}},\,\wbX{g_1}, \dots, \wbX{g_{n-k_2}},
\end{equation}
which generate the so-called stabilizer of the CSS code, 
commute with each other, so that
we have simultaneous eigenspaces of these operators.
Specifically, put
\begin{equation}\label{eq:encoded}
\ket{\phi_{xzv}} = \frac{1}{\sqrt{\crd{\CStwo}}} \sum_{w\in \CStwo}
\omega^{z\cdot w} \ket{x+v+w}
\end{equation}
for coset representatives $x,z$ and $v$.
Then, we have
\[ 
\wbZ{h_j} \ket{\phi_{xzv}}= \omega^{x\cdot h_j} \ket{\phi_{xzv}}, 
\quad j=1,\dots,n-k_1
\]
and
\[
\wbX{g_j} \ket{\phi_{xzv}}= \omega^{z\cdot g_j} \ket{\phi_{xzv}}, 
\quad j=1,\dots,n-k_2 .
\]

It can be checked that
$\ket{\phi_{xzv}}$, 
$x\in\myFpower{n}/\CSone$,
$z\in\myFpower{n}/\CStwp$, $v\in\CSone/\CStwo$, form
an orthonormal basis of $\Hgen$.
In words, we have $\dmn^{k_1+k_2-n}$-dimensional subspaces $\cQ_{xz}$
such that $\bigoplus_{x,z} \cQ_{xz}=\Hch^{\tnsr n}$ and
$\cQ_{xz}$ is spanned by orthonormal vectors $\ket{\phi_{xzv}}$, 
$v\in\CSone/\CStwo$, for each pair $(x, z)\in(\myFpower{n}/\CSone)\times
(\myFpower{n}/\CStwp)$.
The subspaces $\cQ_{xz}$, $(x, z)\in(\myFpower{n}/\CSone)\times
(\myFpower{n}/\CStwp)$,
are the simultaneous eigenspaces of the operators in (\ref{eq:stab4CSS}),
and form a CSS code.

In \cite{hamada03s}, we have treated the case when
$\CSone=\CStwp=\Ccl^{\perp}$ with a code $C$. 
In this case, 
$C$ is necessarily self-orthogonal%
\footnote{%
A subspace $\Ccl$ with $\Ccl\subgrp \Ccl^{\perp}$, 
which is equivalent to $\forall x,y\in \Ccl, \, \dpr{x}{y}=0$,
is said to be
self-orthogonal (with respect to the dot product).}
by (\ref{eq:css_cond}).
We will consistently use $k$ to denote 
the logarithm of the dimension of $\cQ_{xz}$, viz.,
\begin{equation}
k=k_1+k_2-n= \log_{\dmn} \dim_{\bC} \cQ_{xz}. \label{eq:k}
\end{equation}

Decoding or recovery operation for a CSS quantum code can be
done as follows.
If we choose a set $\crJ_i$ of coset representatives of $\myFpower{n}/C_i$ ($i=1,2$),
we can construct a recovery operator $\cR$ for $\cQ_{xz}$
so that the code $(\cQ_{xz},\cR)$
is $\Ebe_{\Kof{\crJ_1,\crJ_2}}$-correcting
in the sense of \cite{KnillLaflamme97},  
where $\Kof{\cdot,\cdot}$ is defined by%
\begin{equation} \label{eq:K}
\Kof{\crJ_1,\crJ_2} = \{ [x,z] \mid x\in\crJ_1 \mbox{ and } z\in\crJ_2 \}.
\end{equation}
In fact, $\cQ_{xz}$ is $\Ebe_{\Kof{\crJ_1',\crJ_2'}}$-correcting with
\begin{equation}\label{eq:css_cr}
\crJ_1'=\crJ_1+C_2^{\perp} \quad \mbox{and} \quad \crJ_2'=\crJ_2+C_1^{\perp}.
\end{equation}
This directly follows from the general theory of symplectic
codes~\cite{crss98,gottesman96}, \cite[Proposition~A.2]{hamada03f} 
on noticing that the operators in the Weyl basis
that commute with all of those in (\ref{eq:stab4CSS}) are
$\wbX{u}\wbZ{w}$, $u\in C_1, w\in C_2$ (see also Sections~\ref{ss:gsc} and \ref{ss:qc}).

\section{CSS Codes as Cryptographic Codes \label{ss:css_crypt}}

Schumacher~\cite[Section~V-C]{schumacher96}, 
using the Holevo bound,
argued that if a good quantum channel code is used as a cryptographic code,
the amount of information leakage to the possible eavesdropper is small. 
In this section, we will apply Schumacher's argument to CSS codes.

\subsection{Quantum Codes and Quantum Cryptography \label{ss:principle_sch}}

Suppose we send a $k$-digit secret information
$\rvv+\CStwo \in \CSone/\CStwo$ physically encoded
into the state $\ket{\phi_{\rvx\rvz\rvv}}\in\cQ_{\rvx\rvz}$, 
where we regard $\rvx,\rvz$ as random variables,
and assume $(\rvx,\rvz)$ are randomly chosen
according to some distribution $P_{\rvx\rvz}$.%
\footnote{The probability distribution of a random variable $\sY$ is denoted by
$P_{\sY}$.}
Once the eavesdropper, Eve, has done an eavesdropping, 
namely, a series of measurements,
Eve's measurement results form another random variable, say, $\rve$.
We use the standard symbol $I$ to denote the mutual information 
(in information theory).

According to \cite[Section~V-C]{schumacher96}, 
\begin{equation}\label{eq:Sch1}
I(\rvv;\rve|\rvx=x,\rvz=z) \le \Se{x}{z}
\end{equation}
where $\Se{x}{z}$ is the entropy exchange after the
system suffers a channel noise $\cN$, Eve's attack $\cE$, another
channel noise $\cN'$, and the recovery
operation $\cR=\cR_{xz}$ for $\cQ_{xz}$ at the receiver's end.
Let us denote by $F_{xz}$ the fidelity of the code $(\cQ_{xz},\cR)$
employing the entanglement fidelity $\Fen$~\cite{schumacher96}.
Specifically,
\[
F_{xz}=\Fen\big(\pi_{\cQ_{xz}}, \cR\cN'\cE\cN\big)
\]
where $\pi_{\cQ}$ denotes the normalized projection operator onto $\cQ$,
and $\cB\cA(\rho)=\cB\big(\cA(\rho)\big)$ for two CP maps $\cA$ and $\cB$,
etc.
Then, by the quantum Fano inequality~\cite[Section~VI]{schumacher96}, we have
\begin{equation}\label{eq:Sch2}
\Se{x}{z} \le h(F_{xz}) + (1-F_{xz}) 2 nR
\end{equation}
where $h$ is the binary entropy function and $R=n^{-1}\log_{\dmn}\dim\cQ_{xz}$.
Combining (\ref{eq:Sch1}) and (\ref{eq:Sch2})
and taking the averages
of the end sides, we obtain
\begin{align}
I(\rvv;\rve|\rvx\rvz) \nonumber\\
&\!\!\! \!\!\! \! \le  \Expe h(F_{\rvx\rvz}) + (1-\Expe F_{\rvx\rvz}) 2 nR \nonumber\\
&\!\!\! \!\!\! \! \le  h(\Expe F_{\rvx\rvz}) + (1-\Expe F_{\rvx\rvz}) 2 nR, \label{eq:Sch3}
\end{align}
where $\Expe$ denotes the expectation operator with respect to $(\rvx,\rvz)$.
Hence, if $1-\Expe F_{\rvx\rvz}$ 
goes to zero faster than $1/n$, then $I(\rvv;\rve|\rvx\rvz) \to 0$
as $n\to\infty$. We have seen in \cite{hamada03s} 
that the convergence is, in fact,
exponential for some good CSS codes,
viz., $1-\Expe F_{\rvx\rvz} \le \dmn^{-n E + o(n)}$ with some $E>0$.
This, together with (\ref{eq:Sch3}), implies 
\begin{equation}
I(\rvv;\rve|\rvx\rvz) \le 
2\dmn^{-nE+o(n)} [n(E  +  R) -o(n)], \label{eq:Sch5}
\end{equation}
where we used the upper bound $- 2 t \log t$ for $h(t)$, $0\le t \le1/2$,
which can easily be shown by differentiating $t\log t$
(or by Lemma~2.7 of \cite{CsiszarKoerner}).
Thus, we could safely send a secret data $v+\CStwo$ 
provided we could send
the entangled state $\ket{\phi_{xzv}}$ in (\ref{eq:encoded})
and the noise level of 
the quantum channel including Eve's action were tolerable by the
quantum code.

In the above scheme,
the legitimate sender, Alice, and receiver, Bob, 
should share the random variables 
$\rvx\rvz$, say, by sending them through 
a public channel that is free from tampering of 
malicious parties (but see Section~\ref{ss:side_inf}). In particular, 
we assume Eve can possibly observe $\rvx\rvz$ without tampering them
as in the literature on quantum key distribution.

\subsection{Reduction to Cryptographic Code}

In \cite{hamada03s}, borrowing the idea of \cite{ShorPreskill00},
we have reduced the above cryptographic scheme to the BB84 protocol.
We now explain that this reduction argument also shows that conjugate code pairs (CSS codes) 
can be used as cryptographic codes.

The above scheme is simply summarized as 
`choose $\rvx\rvz=xz$ randomly, and encode
the secret data into the basis $(\ket{\phi_{xzv}})_v$ of the 
quantum code $\cQ_{\rvx\rvz}$.' 
We would encounter several difficulties and drawbacks
in implementing the above scheme in this form.
Among others,
the state $\ket{\phi_{xzv}}$ defined in (\ref{eq:encoded}) is entangled
in general, and therefore the above scheme is hard to implement
with the current technology.
To overcome this problem, we use Shor and Preskill's observation that 
the probabilistic mixture of $\ket{\phi_{xzv}}$ 
with $x,v$ fixed and $z$ chosen uniformly randomly 
over $\myFpower{n}/\CStwo$
is given as
\begin{multline}
\frac{1}{\crd{\CStwo}}\sum_{z}\ket{\phi_{xzv}}\bra{\phi_{xzv}} \\
= \frac{1}{\crd{\CStwo}}\sum_{w\in \CStwo}\ket{w+v+x}\bra{w+v+x},\label{eq:SPmixed}
\end{multline}
which can be prepared as the mixture of states 
$\ket{w+v+x}$ with no entanglement.%
\footnote{%
A proof of (\ref{eq:SPmixed}) is given in Section~\ref{ss:proofSPmixed}.}
Then, clearly, if Alice sends the secret data $v$ encoded into the state 
in (\ref{eq:SPmixed}) with $x$ chosen randomly according to $P_{\rvx}$
(the marginal distribution of $P_{\rvx\rvz}$), the inequalities
deduced in the previous subsection,
in particular, the bound on the information leakage to Eve
in (\ref{eq:Sch5}), remain true.
Thus, we can send secret data safely by the following scheme.

{\em Conjugate-Code-Based Cryptographic Code}.\/ 
Alice sends a $k$-digit secret information
$\rvv+\CStwo \in \CSone/\CStwo$ physically encoded
into the state in (\ref{eq:SPmixed})
where $\rvx=x$ is chosen randomly 
according to some distribution $P_{\rvx}$.

In this case, Alice and Bob should share 
$\rvx=x$.
We remark 
the random variable $\rvz$, the states $\ket{\phi_{xzv}}$ and
the recovery operator $\cR_{xz}$
are fictitious in that they do not appear
in the above reduced cryptographic code,
but they proved useful for demonstrating the security.
These need only exist in theory, and need not be implemented in practice.

Up to now, we have fixed our attention on proving the security.
However, we should ensure reliable transmission. Namely,
the probability of disagreement between Alice's data $\rvv$ and Bob's data $\rvv'$, 
which should be the result of decoding the cryptographic code, 
must be reasonably small.
For the conjugate-code-based (CSS-code-based) cryptographic code, 
we employ the following decoding
principle.
The receiver performs a decoding algorithm for the coset code 
$x+C_1$ that correct errors in $\crJ_1$, a set of coset representatives
of $\myFpower{n}/C_1$.
The algorithm is the obvious modification of a decoding algorithm for 
the linear code $C_1$.
In the next subsection, we will see that a CSS quantum code with
high fidelity results in a secure and reliable cryptographic code, where a reliable
cryptographic code means that with small decoding error probability.
Note that {\em if $P_\rvx(x)=1$ for some $x$, we do not have to send
$\rvx$}\/ (see the next subsection).


\subsection{Evaluating Fidelity \label{ss:side_inf}}

Note that the underlying CSS codes (before the reduction) 
has the fidelity $\Expe F_{\rvx\rvz}$ which is bounded by 
\begin{equation}\label{eq:Fcss}
1-\Expe F_{\rvx\rvz} \le P_{\cA}(\Kof{\crJ_1',\crJ_2'}\cmple),
\end{equation}
where $J$, $\crJ_1'$ and $\crJ_2'$ are 
as in (\ref{eq:K}) and (\ref{eq:css_cr}).
The right-hand side of (\ref{eq:Fcss}) can be written as
$\Prob \{ \mbox{$\rvxi \notin \crJ_1'$ or $\rvzeta \notin \crJ_2'$} \}$,
and hence we have
$1-\Expe F_{\rvx\rvz} \le 
\Prob \{ \mbox{$\rvxi \notin \crJ_1'$} \} + \Prob \{ \mbox{$\rvzeta \notin \crJ_2'$} \}$,
where $\rvxi$ and $\rvzeta$ are random variables 
such that the distribution of $[\rvxi,\rvzeta]$ is given by $P_{\cA}$.
The equality holds in (\ref{eq:Fcss}) if $\crd{\crJ_1}=\dmn^{n-k_1}$ 
and $\crd{\crJ_2}=\dmn^{n-k_2}$ 
(namely, if they are complete systems of coset representatives).
This follows from that the code is $N_{\Kof{\crJ_1',\crJ_2'}}$-correcting
and that the fidelity of an $N_J$-correcting
symplectic code is $1-P_{\cA}(J)$ for a channel 
$\cA: \Bop(\Hch^{\tnsr n}) \to \Bop(\Hch^{\tnsr n})$, where
the probability distribution $P_{\cA}$ is associated with $\cA$ in the
manner described in \cite{hamada03s,hamada03f} (see also Section~\ref{ss:gsc}).
In the present context, $\cA=\cN'\cE\cN$.
An important fact is that 
the right-hand side of (\ref{eq:Fcss}) is smaller than
$\Prob\{ \rvxi\notin \crJ_1' \}$, 
which is the decoding error probability
when the quotient code $C_1/C_2^{\perp}$ is used as a conjugate-code-based 
cryptographic code. Hence, by bounding the fidelity of the underlying CSS quantum codes, 
we automatically obtain
bounds on the security and reliability simultaneously.

There are subtleties on (\ref{eq:Fcss}).
This fidelity bound is true for a general TPCP map 
$\cA: \Bop(\Hch^{\tnsr n}) \to \Bop(\Hch^{\tnsr n})$ if $P_{\rvx\rvz}$ is uniform.
This is because (\ref{eq:Fcss}) is based on Corollary~4 to Theorem~3
in \cite{hamada03f} or the alternative reasoning in 
\cite[Appendix~A]{hamada03s} and any of these assumes 
the distribution of the syndrome
$(\rvx,\rvz)$ is uniform.
A desirable situation in the reduced cryptographic code
is that the entropy of $P_{\rvx}$ is small.
In particular, if $P_{\rvx}(x)=1$ for some $x$, or (\ref{eq:Fcss}) is true 
for such a random variable $\rvx$ for other reasons,
we do not need the public channel to send $\rvx$. 
This is possible if the map $\cA=\cN'\cE\cN$ is
known to the legitimate participants of the protocol as explained below.

The history of information theory suggests
it would be reasonable to treat first the tractable case 
where $\cE$ is known to the legitimate participants (and $\cA=\cE^{\tnsr n}$)
to pursue the fundamentals of the issue of transmitting private data
(cf.\ \cite{devetak03,CaiWinterY04}). 
\if
the limit of .
since we are still at early stages 
on this issue of transmitting private data
(assuming such assumptions would be pedagogical even the mature). 
\fi
Then, we can interpret the above argument as indicating the existence of 
a good cryptographic code (a kind of random coding proof).
Namely, we can single out the best index $\hat{x}$ such that
$\Expe_{\rvz} F_{\hat{x}\rvz} \ge \Expe_{\rvx\rvz} F_{\rvx\rvz}$,
where $\Expe_{\rvy}$ denotes the expectation operator with respect
to a random variable $\sY$.
Replacing the original random variable $\rvx$ with that whose probability
concentrates on $\hat{x}$, we have a protocol that does not require 
transmission of information $\rvx$ through an auxiliary public channel.

Still, it would be desirable to remove the assumption that 
the legitimate participants
know $\cA$. That is, universal codes that do not depend on the channel
characteristics are desirable.
Regarding this issue, we make a small step forward.
It seems difficult to construct universal cryptographic codes
without transmission of auxiliary information for the completely general class
of channels. 
However, this is possible with our conjugate-code-based cryptographic code
if the class of $\cA$ is restricted to those
such that $\Expe_{\rvz} F_{x\rvz}$ does not depend on $x$.
This situation occurs if, e.g., $\cA$ has the form
$\cA: \rho \mapsto \sum_{u,w\in\myFpower{n}} P(u,w) X^uZ^w \rho (X^uZ^w)^{\dagger}$
with a probability distribution $P$ on $(\myFpower{n})^2$.
This condition is equivalent to that $\cA$ is `Weyl-covariant': $\cN_x\cA=\cA\cN_x$, where $\cN_x:\rho\mapsto N_x \rho N_x^{\dagger}$, $x\in\cX^n$ (see, e.g., \cite[Section~2.5]{hamada03f}).
More generally, the situation occurs if $\cA$ has the property
\[
\cA(X^u \rho X^{-u} ) = X^u \cA( \rho ) X^{-u}
\]
for any $\rho\in\Bop(\Hch^{\tnsr n})$ and $u\in\myFpower{n}$.
This condition is equivalent to 
\[
\bra{l-u}\cA(\ket{i-u}\bra{j-u})\ket{m-u} = \bra{l}\cA(\ket{i}\bra{j})\ket{m}
\]
for any $i,j,l,m,u\in\myFpower{n}$, which reads `the channel looks the same 
if we translate the basis $(\ket{i})_i$ to $(\ket{i-u})_i$.'

\section{General Symplectic Codes \label{ss:gsc}}

In this and next sections, the order $\dmn$ of the finite field
$\myF$ is not necessarily a prime.
In this section,
we digress to explain how general symplectic codes are defined
and how CSS codes are obtained from the general definition.
The $2n$-dimensional linear space $\myFpower{2n}$ over $\myF$ 
equipped with the standard symplectic form
\begin{gather*}
\syp{(x_1,z_1,\dots,x_n,z_n)}{(x'_1,z'_1,\dots,x'_n,z'_n)}\\
 =  \sum_i x_iz'_i - z_i x'_i
\end{gather*}
which has already appeared in (\ref{eq:syp_blf}),
plays a crucial role in algebraic QECCs. 
We can define the dual $L^{\perpsyp}$ of $L$ by 
$L^{\perpsyp}=\{ y \in \myFpower{2n} \mid \forall x \in L, \syp{x}{y}= 0 \}$.
Let us call a subspace $L$ with $L^{\perpsyp} \subgrp L$
an $\sypnoarg$-dual-containing code
or a {\em dual-containing code}\/ (with respect to the symplectic form $\sypnoarg$).
Then, we have a quantum code whose performance is closely related
to that of the classical code $L$. The code is called
a {\em symplectic (quantum) code}\/ with parity check set
$(\varg_1, \dots , \varg_{n-k})$, where $\varg_1, \dots , \varg_{n-k}\in\myFpower{2n}$ 
form a basis of $L^{\perpsyp}$, or a symplectic code with stabilizer 
$N_{L^{\perpsyp}}$.
Here, $N: u \mapsto N_u$
is Weyl's projective representation\mycite{weyl28} of $\myFpower{2n}$
(the same as in Section~\ref{ss:css}).

Suppose $\sA_{n,k}$ is the ensemble of $[2n,n+k]$ $\sypnoarg$-dual-containing codes over 
$\myF$. 
We can regard them
$[n,(n+k)/2]$ additive codes over $\cY=\myFpower{2}$ if we pair up the coordinates of any word $(x_1,z_1,\dots,x_n,z_n)$
to have $((x_1,z_1),\dots, (x_n,z_n))\in\cY^n$. 
We can associate with an $[n,(n+k)/2]$ $\sypnoarg$-dual-containing code a set of
$d^k$-dimensional subspaces of $\sH^{\tnsr n}$,
which can be used for quantum error correction~\cite{crss97,crss98,gottesman96}.
Namely, we have the next lemma,
which is a slight reformulation 
of the original one\mycite{crss97,crss98}.
\begin{lemma} \label{lem:psc}
Suppose a subspace $L \in \sA_{n,k}$ and
a set $\crI$ of representatives of cosets of $L$ in $\myFpower{2n}$ are given.
Then, we have a $\dmn^k$-dimensional subspace of $\sH^{\tnsr n}$
that works as an $N_{\tilde{\crI}}$-correcting code with a suitable recovery
operator,
where $\tilde{\crI}=\crI+L^{\perpsyp}= \{ x+y \mid x \in \crI, y
\in L^{\perpsyp} \}$.
\end{lemma}

For a proof, 
see \Cite{crss98} or, e.g., \Cite{AshikhminKnill00,hamada03f}. 
Roughly speaking, given a set of operators $\soE$,
a quantum code being $\soE$-correcting or a code corrects `errors' in $\soE$
means that it recovers
any state in the code subspace perfectly
after the state suffers `errors' belonging to $\soE$\mycite{KnillLaflamme97}.
The precise definition of $\soE$-correcting is not requisite for 
evaluating the performance of quantum codes.
Indeed, the next fact is enough to treat symplectic codes~\cite{hamada03f}: 
If we properly define the performance
measure 
of symplectic codes, it equals the probability 
$P_{\cA}(\tilde{\crI})$.
The performance measure is the entanglement fidelity averaged over the whole
syndromes, which was already used in Section~\ref{ss:css_crypt}.

A CSS code is a symplectic code with stabilizer $N_{L^{\perpsyp}}$
such that $L^{\perpsyp}$ has the form
$L^{\perpsyp}=\{ [u,w] \mid u \in C_2^{\perp},
w\in C_1^{\perp} \}$ with some $C_1$ and $C_2$.
In this case, $L= 
\{ [u,w] \mid u\in C_1, w\in C_2 \}$, so that
$L^{\perpsyp} \subgrp L$ can be written as
$C_2^{\perp} \subgrp C_1$, the requirement we have posed.

\section{Quotient Codes \label{ss:qc}}

Now, we turn to the realm of finite fields or 
algebraic coding theory. 
In \cite{hamada05qc}, the notion of quotient codes was introduced to
explain QECCs.
The aim of \cite{hamada05qc} was to exhibit the essence,
at least, for algebraic coding theorists,
of algebraic quantum coding.
A {\em quotient code}\/ of length $n$ over $\myF$ is an additive quotient group
$C/B$ with $B\le C\le\myFpower{n}$.
In the scenario of quotient codes in \cite{hamada05qc}, 
the sender encodes a message into a member $c$ of $C/B$, 
chooses a word in $c$
according to some probability 
distribution on $c$, and then sends it through the channel.
Clearly, if $C$ is a $J$-correcting in the ordinary sense,
$C/B$ is $(J+B)$-correcting (since adding a word in $B$ to a code-coset
does not change it).
A conjugate-code-based cryptographic code effectively
means a quotient code in this scenario.
A conjugate-code-based cryptographic code
may be said to be an
error-correcting code that can protect information from eavesdroppers,
and hence may be called a cryptographic error-correcting code.

Lemma~\ref{lem:psc} may read that
if $L$ is a dual-containing code with respect to $\sypnoarg$,
and the quotient code $L/L^{\perpsyp}$ is $\tilde{\crI}$-correcting,
then the corresponding symplectic quantum code is 
$N_{\tilde{\crI}}$-correcting.
Turning our attention to 
the CSS code specified as above with $C_1,C_2$,
the quotient code $L/L^{\perpsyp}$ has the form
$C_1/C_2^{\perp} \oplus C_2/C_1^{\perp}$.
Thus, the CSS code $\cQ_{xz}$ in Section~\ref{ss:css} is
$\Ebe_{\Kof{\crJ_1',\crJ_2'}}$-correcting with
$\crJ_1'=\crJ_1+C_2^{\perp}$ and $\crJ_2'=\crJ_2+C_1^{\perp}$.
In particular, the CSS code has large fidelity if
both $C_1/C_2^{\perp}$ and $C_2/C_1^{\perp}$ 
have small decoding error probabilities.
This is the ground
where the goal described in Section~\ref{ss:cc} stems from.
It might be said 
that the structure of quotient codes were inherent
in quantum error-correcting codes and CSS-code-based cryptographic codes.

\section{Remarks \label{ss:rmk}}

\subsection{Model of Eavesdropping}

A measurement is modeled as
a completely positive (CP) instrument whose measurement result
belongs to a finite or countable set (e.g., \cite{HolevoLN,kraus71,hellwig95,kraus,preskillLNbook}). The specific model employed in this work is the
same as in \cite{hamada03s} and as follows.

We assume a TPCP map $\cA: \Bop(\sH^{\tnsr\varn})\to\Bop(\sH^{\tnsr\varn})$ 
represents the whole action of Eve (plus the other environment).
This means that there exists a decomposition (CP instrument) $\{ \cA_i \}_i$
such that $\cA=\sum_{i} \cA_i$, where $\cA_i$ are 
trace-nonincreasing CP maps, and when the initial state of the system of
the whole sent digits is $\rho$,
Eve obtains data $\rve=i$ with probability $\trace \cA_i(\rho)$
leaving the system in state $\cA_i(\rho) /\trace \cA_i(\rho)$.
Here, the decomposition may depend on the other random variables available to Eve.

A minor comment follows.
Let the random variable $\rve'$
denotes Eve's measurement result on the whole sent digits.
Then, the random variable $\rve$ above mentioned
has more information than $\rve'$ since $\rve$ includes the data 
relevant to the other environment.
However, 
there is no harm in considering $\rve$ as Eve's data for the purpose
of proving the security.

\subsection{Related Information Theoretic Problems}

In \cite{devetak03,CaiWinterY04}, information theoretic 
problems related to ours are treated. These and 
the present work or \cite{hamada03s}
share the goal of secure transmission
of private data, but their specific purpose 
in \cite{devetak03,CaiWinterY04} is to establish coding theorems 
on the best asymptotically achievable rates.
Our codes are linear codes while theirs lack such a helpful structure and
are hard to conceive aimed at practical use.

The quantum theoretical models treated in the literature above mentioned
can be regarded as generalizations of that of \cite{wyner75}.
What are called conjugate-code-based cryptographic codes in the present work
essentially fall in the class of coding systems in \cite{wyner75}.

\subsection{Wiesner's Conjugate Coding}

The term `conjugate coding' appeared 
in the pioneering work on quantum cryptography~\cite{Wiesner83}, 
where the idea of
encoding secret information into quantum states, more specifically,
into conjugate bases, was proposed.
This idea is still alive in CSS-code-based cryptographic codes or QKD schemes. 
However, this is a problem of modulation in the
language of communication engineers. Thus, our meaning of `conjugate'
is different from, though related to, that of \cite{Wiesner83}.

\subsection{QKD Protocol \label{ss:bb84}}

The BB84 QKD protocol as treated in \cite{mayers01acm,ShorPreskill00}
or its variants is, roughly speaking,
the CSS-code-based cryptographic code plus a scheme for estimating the
noise level, where the noise includes the effect of eavesdropping.
Mainly due to the scheme for noise estimation, the protocol needs public
communication. We have used the dichotomy of cryptographic codes and estimation schemes
in analysis of the QKD protocol~\cite{hamada03s},
and have focused more on cryptographic codes in the present work.

\subsection{Non-prime Alphabet \label{ss:npa}}

Let $\dmn=p^m$ with $p$ prime. We have assumed $m=1$ in 
Sections~\ref{ss:css} and \ref{ss:css_crypt}. When $m>1$,
a conjugate code pair $(C_1,C_2)$ over $\myF$ is still useful for
quantum coding and cryptography.
This is because elements of $\myF$ can be
expanded into $\myFnoarg_p^m$ using dual bases
in such a way that ${\rm Tr}_{\myF/\myFnoarg_p}\, x y = \sum_{i} x_i y_i$,
where $(x_1,\dots,x_m)$ is the representation of $x$ with respect to
one basis and $(y_1,\dots,y_m)$ is that of $y$ with respect to the dual~\cite{LidlNied}. Applying these representations to $(C_1,C_2)$, 
we obtain a conjugate code pair over $\myFnoarg_p$. 
This follows easily from \cite[Theorem~1]{KasamiLin88}, or \cite[Theorem~1]{hamada06ccc}.

\subsection{Proof of (\protect\ref{eq:SPmixed}) \label{ss:proofSPmixed}}


The left-hand side can be written as
\[
\frac{1}{\crd{C_2^{\perp}}^2} \sum_{w,w'\in C_2^{\perp}}
\sum_{z} 
\omega^{z \cdot (w-w')} \ket{x+v+w} \bra{x+v+w'}
\]
and we see $\sum_{z}\omega^{z \cdot (w-w')}$ vanishes whenever $w\ne w'$.%
\footnote{%
This follows by an easy direct calculation,
but may be seen as a basic property of characters (e.g., \cite{vanLint3rd}):
the map $f:\, z \mapsto \omega^{z \cdot (w-w')}$ is a character,
and $f(z) \ne 1$ for some $z$ if $w\ne w'$.}
Hence, we have (\ref{eq:SPmixed}).

\subsection{Other Comments}
We take this opportunity to make corrections to related works of 
the present author~\cite{hamada03s,hamada03t,hamada05qc}.
(a) Ref.~\cite{hamada03s}: On p.~8313, line~5, `$\tilde{\Gamma}_n=\Gamma_n + 1^n$' should read
`$\tilde{\Gamma}_n=\Gamma_n + \{ 0^n, 1^n \}$'.
(b) Ref.~\cite{hamada05qc}: On p.~453, right column, 5th line from the bottom, `basis of $L$'
should read `basis of $L^{\perpsyp}$'.
(c) Ref.~\cite{hamada05qc}: On p.~453, right column, 4th line from the bottom, `$N_{L}$'
should read `$N_{L^{\perpsyp}}$'.
(d) Ref.~\cite{hamada03t}: On p.~6, right column, line 16, the period should be removed, and
`With' in the subsequent line should be decapitalized.

\section{Summary and Concluding Remarks\label{ss:cnc}}

Conjugate codes were introduced
without referring to Hilbert spaces so as to be more accessible to 
algebraic coding theorists.
The bridge between the coding theorists' universe, 
the vector space over a finite field, and quantum mechanical worlds
that are represented by Hilbert spaces 
is Weyl's projective representation $N$ of $\myFpower{2n}\simeq \cX^n$,
$N: \cX^n \ni x \mapsto N_x$.
Applicability of conjugate codes to cryptography was argued.
A class of good conjugate code pairs will be given 
in future works~\cite{hamada06ccc,hamada06md,hamada06itw}.

\section*{Acknowledgment}
The author wishes to thank O.~Hirota, Professor of Tamagawa University,
for encouragement.

\end{document}